\documentclass[3p,times,twocolumn]{elsarticle}

\usepackage{ecrc}

\volume{00}

\firstpage{1}

\journalname{Nuclear Physics B Proceedings Supplement}

\runauth{}

\jid{nuphbp}

\jnltitlelogo{Nuclear Physics B Proceedings Supplement}

\usepackage{amssymb}

\usepackage[figuresright]{rotating}

\begin{document}

\begin{frontmatter}



\dochead{}

\title{CTA: the future of ground-based $\gamma$-ray astrophysics}


\author{Massimo Persic (for the CTA Consortium)}

\address{INAF and INFN, Trieste, Italy}
\address{\tt persic@oats.inaf.it}

\begin{abstract}
Very high energy (VHE; $E \geq 100$\,GeV) $\gamma$-rays provide a unique probe into the 
non-thermal processes in the universe. The ground-based Imaging Air Cherenkov telescopes 
(IACTs) for detecting VHE\,$\gamma$-rays have been perfected, so a relatively fast and 
inexpensive assembly of IACTs is now possible. Next generation instruments will have a 
sensitivity $\sim$10 times better than current facilities, and will extend the accessible 
$\gamma$-ray bandwidth at both energy ends (down to 30\,GeV and up to 300\,TeV) with improved 
angular and energy resolutions. Some key physics drivers, that are discussed here, suit 
specific features of the upcoming IACT facility, the Cherenkov Telescope Array (CTA). The 
resulting technical solutions chosen for CTA, and the current status of the project, are also 
outlined.
\end{abstract}


\end{frontmatter}


\section{Introduction}

VHE\,$\gamma$-rays are produced in nonthermal processes in the universe, namely in Galactic 
objects like supernova remnants (SNR), pulsars, pulsar-wind nebulae (PWNe), giant molecular 
clouds (GMC), or binary systems containing compact objects, and extragalactic sources such as 
active galactic nuclei (AGN; mostly blazars and radio-galaxies) and starburst galaxies (SBG). 
Galaxy clusters and $\gamma$-ray bursts (GRB) are also potential, although not yet discovered, 
extragalactic sources of VHE\,$\gamma$-rays. Furthermore, $\gamma$-ray astronomy can be used 
to search for the annihilation of dark matter (DM) particles and to study the transparency and 
star-formation history of the universe; fundamental-physics searches, e.g. for violations of 
the Lorentz invariance, can also be performed. For a review of VHE astrophysics, see Hinton \& 
Hofmann (2009).

Upon reaching the Earth's atmosphere, VHE\, photons or hadrons interact with atmospheric nuclei and 
generate electromagnetic showers that extend over ${\cal O}(10)$\,km in length and $\approx$250\,m 
in width. At VHE, the shower particles are stopped high in the atmosphere. Most secondary shower 
particles, $e^-$ and $e^+$ in the shower core, move with ultra-relativistic speed and cause the 
emission of Cherenkov light. IACTs reflect the Cherenkov light onto multi-pixel cameras that record 
the shower images. The geometry, size, and shape of the recorded shower image retain memory of the 
arrival direction, energy, and nature (photon or hadron) of the incoming 'event'.

\section{Standing issues of ground-based $\gamma$-ray astronomy}

Despite their achievements, current IACTs are affected by limitations that have prevented us 
from reaching a full solution to several issues in VHE astrophysics. 

Current instruments are sensitive at the level of $\approx 10^{-2}$\,C.U. (Crab units
\footnote{ The differential photon flux of the Crab nebula between 60\,GeV and 9\,TeV is $f(E_\gamma) = (6.0 \pm 0.2) \times 10^{-10} x^{a+b\,{\rm log}\,x}$ 
	   TeV$^{-1}$\,cm$^{-2}$\,s$^ {-1}$ with $x \equiv E_\gamma/(300\,{\rm GeV})$, $a = -2.31 \pm 0.06$, $b = -0.26 \pm 0.07$ (Albert et al. 2008).}   
) in the $\sim$0.05-50\,TeV energy range ($5\sigma$ excess events in 50hr observation). At low 
energies, limitations arise due to the background from atmospheric hadronic (and electron) 
showers. At high energies, flux and spectral reconstruction are limited by biases and statistical 
uncertainties. In short, the IACT-reconstructed photon fluxes and energies suffer from the lack 
of a {\it calibrated} cosmic $\gamma$-ray source. Typically, IACTs have field-of-views (FOV) of 
$\sim$3-5\,deg and angular resolutions of several arcmin. 

Future instruments will overcome these limitations, by: 
{\it (i)} improving energy and angular resolution; 
{\it (ii)} increasing sensitivity in the core range (100\,GeV-50\,TeV); 
{\it (iii)} pushing the low-energy threshold down to $E_\gamma <50$\,GeV; and 
{\it (iv)} acquiring sensitivity at $E_\gamma >50$\,TeV. 

The Cherenkov Telescope Array (CTA) is the upcoming next-generation IACT that will implement the 
improvements outlined above. Let us now briefly comment on each of the expected improvements.

\subsection{ Increased temporal and angular resolution }

By reducing observation time (for a given source flux) the increased sensitivity of CTA 
will lead to an improved temporal resolution w.r.t. current instruments: the latter are 
sensitive to flux variations on a few minutes timescales, whereas CTA aims at a sub-minute 
capability. CTA also aims at $\sim$0.05$^o$ angular resolution (at $>$1\,TeV), i.e. better 
by a factor of $\sim$2 than those achieved by current IACTs. 

This improvement will help to (e.g.): {\it (i)} pinpoint the location of $\gamma$-ray 
flares, and hence understand the underlying acceleration mechanisms; {\it (ii)} study 
spatially extended sources (e.g., SNRs, PWNe, GMC), both morphologically and spectrally; 
{\it (iii)} reduce source confusion in the crowded Galactic plane region.

\subsection{ Core energy region (100\,GeV$<$$E_\gamma$$<$50\,TeV) }

This is the energy band where most of the action in VHE astrophysics occurs, because 
the sensitivity of IACTs is highest in this energy band hence any spectral features 
are best revealed. HESS, and later VERITAS, have led the way here. 

\smallskip

\noindent
{\it Surveys.} CTA is expected to achieve a tenfold sensitivity increase in this core 
energy band as compared to current IACTs, reaching a level of mCU sensitivity ($5\sigma$ 
photon excess in 50\,hr) in this range. This will be extremely important for surveys (see 
Dubus et al. 2013). 
{\it (a)} At least for the Southern array, a very meaningful, complete flux-limited sample 
of Galactic sources will be obtained if a Galactic plane survey will be performed: for 
example, a HESS-like ($|l| \leq 60^o$, $|b| \leq 2^o$) Galactic plave survey down to a 
sensitivity of 3\,mCU can be completed in 250\,hr, perhaps leading to the detection of 
hundreds of sources. This will permit to build a cumulative luminosity function (LF) of 
Galactic accelerators -- and, when all the CTA-detected sources will have multi-frequency 
identifications, of the separate LFs of the different source classes. 
{\it (b)} Besides and beyond the Milky Way, a 1/4th of the sky may get scanned in 
370\,hr down to a sensitivity of 20 mCU using the wide-FOV mode, or alternatively the whole 
sky may get scanned in 100\,hr using the divergent-pointing mode with no substantial loss 
of sensitivity: either way, some tens of new blazars would most likely be discovered. CTA's 
all-sky survey would nicely complement LAT's all-sky survey at lower energies. 
\smallskip

\noindent
{\it Spectral features.} 
The increased sensitivity in the mid-energy region will permit deeper spectral studies of 
individual accelerators. 

Spectral curvature in the TeV region is crucial for understanding the emission physics of 
extragalactic sources (e.g., Sol et al. 2013; Reimer \& B\"ottcher 2013). The latter's 
emitted radiation interacts with the lower energy photons of the Extragalactic Background 
Light (EBL) on their way to the Earth. The EBL is the integrated light from all the stars 
that have ever formed, and spans the IR-UV range. The interaction of VHE\,$\gamma$-rays 
(with energy $E_\gamma$), emitted by faraway sources, with the intervening EBL photons 
(with energy $\epsilon_\gamma$) may result in $e^-e^+$ pair production that leads to an 
energy-dependent attenuation of the observed VHE flux. This introduces a fundamental 
ambiguity in the interpretation of the measured VHE blazar spectra: neither the intrinsic 
spectra, nor the EBL, are separately known -- only their combination is. Since the 
$\gamma$-$\gamma$ cross section has a resonance for $E_\gamma \epsilon_\gamma \simeq 10^{12}
$\,eV$^2$, a TeV photon will mostly interact with an eV photon. To measure the EBL with IACTs, 
a method proposed by Mankuzhiyil et al. (2010) can be used. It relies on using simultaneous 
observations of blazars in the optical, X-ray, high-energy (HE: $E>100\,$MeV) $\gamma$-ray 
(e.g., from LAT), and VHE\,$\gamma$-ray bands. For each source, the method involves 
best-fitting the spectral energy distribution (SED) from optical through HE\,$\gamma$-rays 
(the latter being largely unaffected by EBL attenuation out to $z \approx 1$) with a 
Synchrotron Self-Compton (SSC) model. We extrapolate such best-fitting model into the VHE 
regime, and assume it represents the blazar's intrinsic emission. Comparison of measured and 
intrinsic emission leads to a determination of the $\gamma$-$\gamma$ opacity to VHE photons -- 
hence, upon assuming a specific cosmology, we derive the EBL photon number density. In 
principle, this EBL mesurement can be made at several redshifts, so enabling different 
measures of EBL at different $z$. For an extended discussion of how CTA can contribute 
to EBL studies, see Mazin et al. (2013).

In Galactic sources spectral curvatures are most likely intrinsic -- unless local photon 
densities are high enough to cause significant VHE\,$\gamma$-ray absorption. Any unveiled 
spectral feature will be important to determine the underlying emitting particles spectrum 
and hence the acceleration mechanism. 
\smallskip

\noindent
{\it Galactic cosmic rays (CRs): origin, propagation, confinement.} Owing to improved 
sensitivity, CTA observations of circum-SNR regions may constrain the 
diffusive behavior of CRs through the latter's interaction with the nearby ($\sim$\,kpc) 
ambient gas: if a giant molecular cloud is located in the vicinity of the CR source, the 
expected emission may be detected by CTA even if the cloud is located ${\cal O}(10)$\,kpc 
away from us. After escaping the sources, CRs diffuse in the turbulent magnetic field of 
the galaxy and finally escape into the intergalactic medium. Because of this, galaxies 
and clusters of galaxies are expected to be detectable $\gamma$-ray emitters (see Acero 
et al. 2013). In particular, CTA is expected to detect further starburst galaxies (SBGs), 
in addition to M\,82 and NGC\,253, thanks to its tenfold better than current sensitivity 
that will permit sampling a $\sim$30 times larger volume of the local Universe. CTA 
observations of SBGs are expected to return information on the proton diffusion coefficient, 
$\delta$: in the leaky-box approximation, $\Gamma = \Gamma_{\rm inj}+\delta$ (where $\Gamma_
{\rm inj}$ and $\Gamma$ are the injection and steady-state proton spectral indices. Also, 
CTA will help measuring the spectrum of CR electrons -- a component that although small 
($\small$1\% of the total CR flux) provides important information on the origin and 
propagation of CRs that is not accessible from CR nuclei due to their differing energy 
loss processes. Safe particle identification is required to separate electrons from a vast 
background of nuclei: this identification is based upon calorimetry information (for CTA 
the calorimeter will be the atmosphere itself), e.g. energy losses and shower development: 
thickness of calorimeter and soundness of simulations are crucial to rejecting hadrons and 
to estimating amospheric secondary electrons, and hence the systematic uncertainties 
affecting the CR electron spectrum (see Picozza \& Boezio 2013).

\subsection{ Low-energy ($E_\gamma$$<$100\,GeV) physics }

MAGIC has led the access to the sub-100\,GeV domain. Building on that experience, CTA will 
further lower the low-energy threshold to better tackle some key astrophysics issues. 
\smallskip

\noindent
{\it Calibration of CTA.} One major reason to push the low-energy reach of CTA to lower 
energies is the wish to enlarge the spectral overlap between CTA and space-borne $\gamma$-ray 
telescopes operating in the HE regime by {\it direct} detection of $\gamma$-ray photons, such 
as LAT. CTA's reconstructed $\gamma$-ray photons' fluxes and energies can be compared 
with LAT's measured ones, hence any biases arising from the reconstruction process may become 
known and corrected for (see Funk \& Hinton 2013).
\smallskip

\noindent
{\it SNRs.} A key issue in understanding $\gamma$-ray of emission of SNR is whether it is 
hadronic or leptonic in nature (see Berezhko \& V\"olk 2006 and Aharonian et al. 2006 on 
SNR\,J1713.7-3946J) -- the former referring to the $\pi^0$-decay emission following the 
interaction between an energetic proton and an ambient proton, whereas the latter refers 
to emission from lower energy photons that get upscattered off energetic electrons and gain 
a factor $\approx \gamma_e^2$ in energy. Over a limited bandwidth, the two emissions look 
similar with current data: but over a bandwidth more extended to low energies, the flatter 
or more cusped profiles characterizing the two emission spectra should become apparent, so 
clarifying the nature of the emission (purely hadronic, purely leptonic, or a combination 
of both). The issue is clearly very relevant also in view of the supposed link between SNRs 
and the origin of Galactic CRs. 
\smallskip

\noindent
{\it Pulsars.} Data from LAT have shown that most pulsars have spectral cutoffs in the GeV (and 
even sub-GeV) range, so detection of pulsar emission may require reaching down to few tens of 
GeV. However, if the recently discovered pulsed emission from the Crab pulsar up to 400\,GeV 
(Aliu et al. 2008 and 2011; Aleksic 2012), unexpected on theoretical grounds and well above the 
LAT cutoff energies, is an important spectral component common to many pulsars, then CTA may 
even detect a substantial population of sources for which LAT has established a sub-GeV spectral 
cutoff. Based on the Crab case, the still unsettled issue of pulsar emission will clearly benefit 
from CTA's extension to lower energies and enhancement of core energy sensitivity, as well as 
from its increased temporal resolution, that will permit a detection of the pulsed emission from 
the Crab pulsar in $<$1\,hr, finding the interpulse in the lightcurve up to 1\,TeV (see 
de\,O$\tilde{\rm n}$a-Wilhelmi et al. 2013; De los Reyes \& Rudak 2013). 
\smallskip

\noindent
{\it PWNe.} A related class of sources are PWNe (see Reynolds et al. 2012 for a review). 
These are young SNR whose pulsar's instantaneous loss of rotational kinetic energy, $\dot{E}$$
\sim$$10^{32}$--$10^{39}$\, erg\,s$^{-1}$, mostly goes into an outflowing relativistic 
magnetized particle wind rather than into the pulsed electromagnetic radiation observed from 
the pulsar: due to external pressure, this wind abruptly decelerates at a termination shock, 
beyond which the pulsar wind thermalizes in pitch angle and radiates synchrotron emission -- 
so resulting in a PWN. The evolution of a PWN can be summarized as {\it (i)} expansion of the 
PWN into the unshocked ejecta, {\it (ii)} interaction of the PWN with the surrounding SNR, 
{\it (iii)} supersonic motion of the powering pulsar through the SNR. The best know case of a 
PWN is the Crab Nebula, powered by the central young pulsar B0531+21: it emits synchrotron 
radiation across the whole electromagnetic spectrum, and its extent grows from X-rays through 
optical to radio as a consequence of the increasing electron lifetime, which in the radio 
exceeds the age of the source ($\sim$1000 yr). A SNR is powered instantly by the explosion, 
whereas a PWN is powered gradually through a relativistic flow of $e^+e^-$ pairs. As a 
consequence, the radio morphologies of SNR and PWN differ, the former being a limb brightened 
shells of synchrotron emission, the latter being an amorphous to symmetric center-filled 
synchrotron nebulae, brightest at the pulsar position. 

TeV observations of PWNe revealed them to be the most efficient type of source in producing 
$\gamma$-rays through the inverse-Compton (IC) mechanism. This IC peak contains information 
on the total energy released by the pulsar in its lifetime, because of the cooling time of 
TeV electrons being longer than that of keV-emitting electrons. PWNe thus act as calorimeters 
at VHE and CTA observations will permit an investigation of the time evolution of these systems 
-- including a better understanding of the still unknown mechanism of how the pulsar releases 
its rotational energy to a kinetic wind with Lorentz factors of $\approx 10^6$ at the 
termination shock. As far as the temporal and morphological evolution is concerned, two 
main types seem to be emerging: the younger plerions where VHE and X-ray morphologies match, 
and older systems where the pulsar powering the TeV emission is seen offset w.r.t. the 
latter. The evolution of the SNR shock wave into an inhomogeneous interstellar medium and/or 
the high velocity of the pulsar in the reference frame of the SNR barycenter, together with 
a low ($\approx 5\,\mu$G) magnetic field may explain these conspicuous offsets as being the 
relic of the past history of the pulsar wind inside the SNR. PWNe are the largest and most 
efficient CR-accelerating population of VHE sources, so a deeper understanding of their 
physics and evolution is a key CTA goal (see de O$\tilde{\rm n}$a-Wilhelmi et al. 2013). 
\smallskip

\noindent
{\it Binaries.} Binary stars detected in $\gamma$-rays have been grouped in different classes 
depending on the nature of the binary components or the origin of the particle acceleration. 
The latter can be interaction of the winds of either a pulsar and a massive star or two massive 
stars, accretion onto a compact object and jet formation, and interaction of a relativistic 
outflow with the external medium. Studying these objects requires observations at different 
temporal and spatial scales. Given the possible competition of leptonic and hadronic processes, 
the low-energy extension of CTA will be able to probe the physical processes underlying the 
$\gamma$-ray emission of binaries with high spectral, temporal, and spatial resolution (see 
Paredes et al. 2013).
\smallskip

\noindent
{\it AGNs, GRBs.} Reaching down in energy means enlarging the $\gamma$-ray horizon. The latter 
is defined as the redshift at which the EBL-related optical depth to VHE photons is unity, i.e. 
$\tau_{\gamma\gamma}(E_\gamma, z)=1$. For $E_\gamma < 50$\,GeV the photon-photon interaction rate 
is relatively low, due to the decreasing density of the ambient UV photons: these low-energy 
$\gamma$-rays can then travel large distances unscathed. Consequently, high-luminosity (because 
beamed) sources like AGN and GRB can be detectedout to large $z$, unlike what happens at higher 
$E_\gamma$. The $<$50\,GeV spectra of blazars, being basically unaffected by the EBL (in the 
relevant range of redshift, out to $z \approx 1$), can be used -- in conjunction with simultaneous 
multi-frequency data -- to reconstruct the emitted SED, and then to solve the SSC model. This 
ensures that even in lack of post-{\it Fermi} space-borne HE\,$\gamma$-ray telescopes, CTA will 
make it possible to measure the portion of the blazar SED at or around the Compton peak and hence 
to permit a full SSC modelization of blazar spectra. As for GRBs, no event was so far detected by 
an IACT, so current models are unconstrained by VHE data -- only a handful of $>$100\,GeV photons 
have been captured by LAT. So CTA may provide a unique instrument, of huge effective 
area ($\approx 10^4$\,m$^2$ (to be compared with LAT's $\approx 1$\,m$^2$ area) in an energy band 
where the Universe is still transparent to $\gamma$-rays, to significantly enlrage the number of 
TeV AGNs and to finally detect GRBs (see Reimer \& B\"ottcher 2013; Sol et al. 2013; Takahashi et 
al. 2013).
\smallskip

\noindent 
{\it Unidentified Fermi Objects.} CTA's low-energy sensitivity will be useful to complement 
the data obtained by LAT at its high-energy end with CTA data of higher statistical 
significance. This will help understanding the nature of the tens of yet unidentified LAT 
sources (see Bednarek 2013).
\smallskip

\noindent
{\it DM.} DM particle candidates should be weakly interacting with ordinary matter 
(and hence neutral). The theoretically favored ones, which are heavier than the proton, are 
dubbed weakly interacting massive particles (WIMPs). WIMPs should be long-lived enough to have 
survived from their decoupling from radiation in the early universe into the present epoch. 
Except for the neutrino, which is the only DM particle known to exist within the Standard Model 
of elementary particles (with a relic background number density of $\sim$50\,cm$^{-3}$ for each 
active neutrino species) but which is too light (sum of the three species of neutrino mass is 
$\approx$1\,eV) to contribute significantly to $\Omega_m$ (e.g., Munoz 2004). 
DM WIMP candidates have been proposed only within 
theoretical frameworks mainly motivated by extensions of the Standard Model of particle 
physics (e.g., the R-parity conserving supersymmetry [SUSY]). Among current WIMP candidates, 
the neutralino, which is the lightest SUSY particle, is the most popular candidate. Its relic 
density is compatible with $\Omega_m$if its mass, $m_\chi$, is in the GeV-TeV range. Its 
self-annihilation leads to a $\pi^0$ that decays into two $\gamma$-ray photons. Worthwhile 
astrophysical targets are those which, seen as point-like, maximize $\rho_{\rm DM}^2/d^2$ and 
have a negligible astrophysical $\gamma$-ray background. Current IACTs and LAT) 
sample, respectively, the high-$m_\chi$ and low-$m_\chi$ regions. Only upper limits exist on 
the quantity $<$$\sigma v_{\rm th}$$>$, i.e. the neutralino self-interaction cross-section 
averaged over the thermal distribution of velocities at their decoupling from radiation: 
these limits come mostly from the Galactic Center, dwarf galaxies, and galaxy clusters. CTA 
will explore a broad mass range, 50\,GeV$ < m_\chi c^2 < $50\,TeV, with unprecedented 
sensitivity: for example, simulations show that some CTA configurations can reach twice the 
sensitivity of HESS for $m_\chi=0.5$\,TeV (see Doro et al. 2013).

\subsection{ High-energy ($E_\gamma$$>$50\,TeV) physics }

There are several cases where high-energy observations with CTA may prove crucial. 
\smallskip

\noindent
{\it Origin of Galactic CRs.} CTA's extension of the spectral window to higher energies than 
current IACTs will allow exploration of hitherto unknown parts of source spectra. A main item 
in this field will be the Origin of Galactic CRs. The deepest SNR observation (J1713.7-3946), 
that stacks three years of HESS data (Aharonian et al. 2007), have yielded a spectrum that 
extends to $\sim$100\,TeV and shows a cutoff at $\sim$30\,TeV. If the emission is leptonic the 
parent electrons have an energy of up to $E_e \approx 100$\,TeV; if the emission is hadronic 
the parent protons have $E_p \approx 200$\,TeV. Whatever the case, the involved particles have 
maximum energies that fall short of just about an order of magnitude in energy of the knee of 
the Galactic CR spectrum (Aharonian et al. 2007). A deeper high-energy observation of this 
(and other) SNR could clarify whether the spectral cutoff seen by HESS in SNR\,J1713.7-3946 is 
intrinsic or is due to insufficient sensitivity. In short, CTA can be used to find the PeVatrons 
that would produce ${\cal O}(10^2)$ TeV photons (see Acero et al. 2013). 
\smallskip

\noindent
{\it Cosmic Infrared Background (CIB).} About half of the EBL is emitted in the IR range (8-1000\,
$\mu$m) and is called the CIB. At far-IR wavelengths ($\lambda > 100\,\mu$m, i.e. $\epsilon_\gamma 
< 1.24 \times 10^{-2}$\,eV) band of the CIB is poorly known due to the rather low instrumental 
angular resolution that strongly limits the depth of galaxy number counts (B\'ethermin \& Dole 2011). 
Therefore blazar observations at $E_\gamma > 80$\,TeV may prove unique in shedding light on this 
relatively unprobed part of the EBL. 
\smallskip

\noindent
{\it New Physics.} Detecting flares from distant AGNs or GRBs at ${\cal O}(10^2)$ TeV energies 
would also increase our prospects for putting firm limits on Lorentz invariance violation. Most 
theories on Quantum Gravity (QG) predict non-conventional light-dispersion relations, of the type 
$c^2p^2 = E^2\,[1 + \xi_1 E/E_{\rm QG} + \xi_2 {\cal O}(E^2/E_{}^2)]$ for $E$$<<$$E_{\rm QG}$ 
(with $E_{\rm QG}$$\sim$$E_{\rm Planck}$$\equiv$$\sqrt{\hbar c/G}$$\approx$$1.22 \times 10^{19}$\,GeV) 
and $\xi$$=$$\pm$$1$ depending on the dynamical framework (e.g., Amelino-Camelia et al. 1998). So, 
$v$$=$$\partial E/\partial p$$\sim$$c\,[1 + \xi_n (E/E_{\rm QG})^n]$ with $n$$=$$1$ or $2$, i.e. 
photons would have energy-dependent velocities. This would mean that the vacuum responds 
differently to the propagation of particles of different 
energies, and the QG space-time fabric would fluctuate on scales $\lambda$$\sim$$\lambda_{\rm 
Planck}$$\approx$$10^{-33}$\,cm on timescales $t$$\approx$$h/E_{\rm QG}$$\approx$$h/E_{\rm Planck}$. 
The arrival time delay for two photons with respective energies $E_{\rm min}$ and $E_{\rm max}$, 
is $\Delta t$$\sim$$\xi [(E_{\rm max}^n-E_{\rm min}^n)/E_{\rm QG}^n]\, d/c$ with $n$$=$$1$ or $2$ if 
the deviation of the photon velocity from $c$ is linear or quadratic in $E$. Stricter lower 
limits on $E_{\rm QG}$ will be obtained in the case of photons of vastly different energies, 
coming from a very distant source, and detected on a very small time delay. Current limits are 
$E_{\rm QG, \,1}$$>$$0.17\,E_{\rm Planck}$ and $E_{\rm QG,\, 2}$$>$$5.25 \times 10^{-9}E_{\rm Planck}$ 
(Abramowski et al. 2011). CTA is expected to lead to tighter lower limits on $E_{\rm QG,\, n}$, 
by virtue of its higher maximum energy ($E_{\rm max}$), time resolution ($\Delta t$), and flux 
sensitivity (that means, for a given source luminosity, a larger $d$). For details see Doro et 
al. (2013).

\section{ CTA: basic concepts }

To meet the physics requirements specified in the previous section with a high technical 
performance within a reasonable budget, the CTA concept is based on few general ideas: 
{\it (a)} use proven IACT technology; 
{\it (b)} increase the array from currently 4-5 telescopes (VERITAS, HESS) to several tens; 
{\it (c)} use telescopes of 3 different sizes, that are best suited to operate in the three 
above mentioned energy bands in a balanced way such that the sensitivity is overall optimized 
at all energies: 
{\it (i)} few ($\approx$4) large-size telescopes (LST) with $\approx$23\,m diameter parabolic
dishes, optimized for $E_\gamma < 200$\,GeV region (based on MAGIC), with a $5^o$ FOV, and 
placed at the center of the array; 
{\it (ii)} several ($\approx$20) medium-size telescopes (MST) with $\approx$11\,m diameter 
Davies-Cotton dishes, optimized for use in the range $0.1-10$\,TeV, with a $8^o$ FOV, and 
placed in a central corona of the array surrounding the LST nucleus: these telescopes (based 
on HESS and VERITAS) will be the core of the array -- and will also veto the LST triggers in 
order to reduce the hadronic background; and 
{\it (iii)} many ($\approx$55) simple-to-build small-size telescopes (SST) with $\approx$4\,m 
or $\approx$7\,m diameter Davies-Cotton or Schwarzschild-Couder dishes, optimized for $E_\gamma 
> 5$\,TeV, with a $10^o$ FOV, and scattered over a large area on ground; 
{\it (d)} distribute telescopes over a large area ($\approx 1 - 10$\,km$^2$) on the ground; 
{\it (e)} provide high automatization liable to remote operation; and 
{\it (f)} run facility as an open observatory, open to the astrophysics community.
In general, a larger number of nuclear LSTs improves the sensitivity at low energies; a 
larger and more scattered number of LSTs improves the sensitivity at high energies. So for 
a given budget, the exact number of telescopes of each type, their size and configuration, 
are being investigated as a function of the array's overall performance (see Bernl\"ohr 
et al. 2013). 

Different CTA operation modes will be possible. In the {\it deep-field} mode all telescopes 
will be pointed to the same sky position to maximize sensitivity. In the {\it divergent-pointing} 
different subsets of telescopes could point to different sky positions -- to make 
simultaneous observations of different sources. In the {\it wide-FOV} mode, e.g., an all-sky 
scan can be performed in a time-efficient way at moderate sensitivity.

Two arrays are planned, in the Southern and Northern hemisphere respectively. The Southern array 
may possibly be the main one, owing to its advantage location in observing the central region of 
the Milky Way. The Northern array will be mostly devoted to extragalactic observations: in this 
case, both the small local pair-production mean free path expected for $E_\gamma>50$\,TeV photons 
($\lambda$$<$$2$\,Mpc, see Coppi \& Aharonian 1997) and the low intrinsic emission expected from 
either steep proton spectra or extreme-KN IC electron scattering, may strongly limit the use of 
SSTs in CTA-North.

\section{ The CTA consortium }

CTA was started as a partnership between the HESS and MAGIC collaborations, but several more 
institutions later joined in. CTA is organized as a consortium that currently comprises $>$900 
scientists and engineers from $>$100 institutions in 26 countries world-wide. 

Since late 2010 (and presumably until late 2013) the Preparatory Phase of CTA has been taking 
care of making technical designs, evaluating sites, optimizing the physics return, estimating 
construction and operation costs, assessing legal/governance/finance schemes, and building 
telescope prototypes. 

The aim is to start deploying CTA in 2014 and to get first full-array data in 2018 -- but partial 
operation could start as early as 2015.

\section{ Concluding remarks }

CTA's promised output will be a mix of guaranteed and discovery science. Technology-wise it is 
a safe extrapolation of proven technologies into a well-predictable performance. It is well 
supported by a large and diverse community, and it's highly ranked by major European and US 
science roadmaps for reasearch infrastructures.

Some of the physics drivers that call for CTA have been described in this paper. The effort to 
build CTA may well be rewarded by a obtaining a full solution to some important problem in 
(astro)physics and cosmology, such as the origin and propagation of Galactic CR, the spectral 
emission and evolution of the EBL, the break of Lorentz invariance, the nature of DM -- by 
having the prime, and likely unchallenged for many years, tool for viewing the non-thermal side 
of the Universe.

\label{}




\nocite{*}
\bibliographystyle{elsarticle-num}
\bibliography{martin}








\end{document}